\documentclass{article}
\usepackage{appendix}
\usepackage{amssymb, amsmath, amsfonts, xcolor, verbatim}
\usepackage{graphicx, subcaption}%
\usepackage[backend=biber, style=apa ,natbib=true, sortcites]{biblatex}
\usepackage{tikz, float}
\usetikzlibrary{calc,fit,arrows}

\usepackage{wrapfig}
\usepackage{lscape}
\usepackage{rotating}
\usepackage{epstopdf}

\immediate\write18{texcount -inc -incbib

-sum main.tex > /tmp/wordcount.tex}

\providecommand{\U}[1]{\protect\rule{.1in}{.1in}}

\usepackage{hyperref}
\newcommand{\footremember}[2]{%
  \footnote{#2}
    \newcounter{#1}
    \setcounter{#1}{\value{footnote}}%
}
\newcommand{\footrecall}[1]{%
    \footnotemark[\value{#1}]%
} 

\title{Building A Bayesian Decision Support System for Evaluating COVID-19 Countermeasure Strategies}
\author{Peter Strong \footremember{complexity}{Centre for Complexity Science, University of Warwick, Coventry CV4 7AL, UK} \\ \href{mailto:P.R.Strong@warwick.ac.uk}{P.R.Strong@warwick.ac.uk} 
\and Aditi Shenvi \footrecall{complexity} \\ \href{mailto:A.Shenvi@warwick.ac.uk}{A.Shenvi@warwick.ac.uk}\and Xuewen Yu \footremember{stats}{Department of Statistics, University of Warwick, Coventry CV4 7AL, UK} \\ \href{mailto:Xuewen.Yu@warwick.ac.uk}{Xuewen.Yu@warwick.ac.uk} \and K.Nadia Papamichail \footnote{Alliance Manchester Business School, The University of Manchester, Booth Street West, Manchester, M15 6PB, UK} \footremember{turing}{The Alan Turing Institute, British Library, 96 Euston Road, London NW1 2DB, UK}  \\ \href{mailto:nadia.papamichail@manchester.ac.uk}{nadia.papamichail@manchester.ac.uk} \and Henry P Wynn \footnote{London School of Economics and Political Science, Houghton Street, London WC2A 2AE, UK}  \footrecall{turing} \\ \href{mailto:H.Wynn@lse.ac.uk}{H.Wynn@lse.ac.uk} \and Jim Q Smith \footrecall{stats} \footrecall{turing}\\ \href{mailto:J.Q.Smith@warwick.ac.uk}{J.Q.Smith@warwick.ac.uk}}

\begin{document}
\maketitle
\begin{abstract}
Decision making in the face of a disaster requires the consideration of several complex factors. In such cases, Bayesian multi-criteria decision analysis provides a framework for decision making. In this paper, we present how to construct a multi-attribute decision support system for choosing between countermeasure strategies, such as lockdowns, designed to mitigate the effects of COVID-19. Such an analysis can evaluate both the short term and long term efficacy of various candidate countermeasures. The expected utility scores of a countermeasure strategy capture the expected impact of the policies on health outcomes and other measures of population well-being. The broad methodologies we use here have been established for some time. However, this application has many novel elements to it: the pervasive uncertainty of the science; the necessary dynamic shifts between regimes within each candidate suite of countermeasures; and the fast moving stochastic development of the underlying threat all present new challenges to this domain. Our methodology is illustrated by demonstrating in a simplified example how the efficacy of various strategies can be formally compared through balancing impacts of countermeasures, not only on the short term (e.g. COVID-19 deaths) but the medium to long term effects on the population (e.g increased poverty). 
\end{abstract}

\section{Introduction}
\label{sec:intro}

Major disasters and incidents such as pandemics, nuclear disasters, volcano eruptions and tsunamis can impact the short-term and long-term well-being and health of people, nature and economies (see e.g. \citet{boyd2014emergency, jorda_singh_taylor_2020, ohtsuru2015nuclear, tamura2000modeling}). Decision-making in such crises involves balancing several complex factors, where each factor's priority may be a subjective judgement \citep{smith2010bayesian}. Such decision-making is further complicated by the decision centre (DC)  -- an individual or a group of people, such as a government committee, who must have the authority to enact the decisions made -- receiving information from multiple sources. Such information is either factual or narrative in nature, and also may be noisy, uncertain and incomplete \citep{bakker2019decision}. Bayesian decision analysis has been shown to enable principled decision-making in such complex, multi-faceted problems \citep{smith2010bayesian, moffat2002bayesian, french1996multi, kurth2017trends}.

This paper focuses on implementing the well-developed principles of Bayesian multi-criteria decision analysis on the ongoing COVID-19 pandemic which first emerged in Wuhan, China in late 2019 \citep{who}. Using the framework presented in this paper, a DC, such as the government, can combine streaming uncertain information they receive about the current and potential effects of the pandemic on the various aspects of society in the form of expected utility scores to decide among a finite set of countermeasure strategies. The study within this paper was undertaken between June and October 2020. 

In March 2020, most countries across Europe, including the UK, imposed lockdowns on regional or national levels in an effort to control the spread of the COVID-19 virus, prevent overburdening their healthcare systems and to buy more time to enable researchers to learn more about the disease \citep{sridhar_2020}. However, lockdowns have been economically damaging, particularly for hard-hit sectors such as aviation, tourism and hospitality \citep{coronavirus_impact}. The UK was in recession in the first two quarters of 2020 with early signs of recovery in the third quarter when lockdown restrictions were relaxed \citep{pwc_report}. As of 18th October 2020, 9.6 million jobs have been furloughed under the UK government's job retention scheme \citep{hmrc_statistics}. Following the relaxation of initial national lockdown measures in summer of 2020, there have been several localised outbreaks of COVID-19 across the UK, resulting in local lockdowns. The government, keen to avoid another national lockdown to protect the economy, has been under scrutiny regarding the efficacy and frequent changes of the measures introduced under local lockdowns \citep{bbc_local_lockdown}. In this ``second wave", the government has again faced a challenging balancing act: managing the public health impacts of COVID-19 on one hand, its economic and social impacts on the other. The research presented in this paper was completed against this backdrop; it reached completion in October 2020, when COVID-19 cases were increasing rapidly and the government was on the brink of announcing a second lockdown.

To compare the health, social and economic impacts of the candidate countermeasure strategies, it is necessary to evaluate them on a comparable scale. Typically, health and social impacts are measured in terms of health-adjusted life-years (QALYs) \citep{zala2020costing, miles2020stay} or wellbeing-adjusted life-years (WELLBYs) \citep{layard2020release, de2020taking} as caused directly by the virus itself or indirectly by the countermeasures implemented to tackle it; the economic impacts are measured in terms of difference in realised GDP and empirical or projected GDP, or sometimes as number of jobs in various sectors, consumer spending growth or business investment growth \citep{pwc_report}. Further, these impacts often move in opposite directions in response to any given countermeasure. Hence, a \textit{common metric} is essential for their comparison \citep{layard2020release}. 

Within a Bayesian decision analysis, this common metric is defined through an \textit{expected utility score} given to each of the health, social and economic attributes. For each attribute, we define a function to convert the actual recorded or estimated measurements (e.g. number of deaths or first order difference in the GDP) into a common metric (here, utilities). Next, by combining these expected utility scores using a specified utility function, we arrive at an expected utility score for each countermeasure strategy. However, the exercise of arriving at an expected utility score for each strategy is complicated by two factors: (1) the specification of the DC's utility function may involve uncertain estimates of future events (e.g. predicted number of COVID-19 deaths under a complete lockdown), and (2) individuals within a DC may be inclined to prioritise the importance of the various impacts in different ways (e.g. some individuals may view the impact on GDP as more important than others). Consequently, there may be no consensus within a DC as to how efficacious each policy may be. 

In this paper, we argue that, in light of the aforementioned challenges, it is essential that any decisions made by policymakers are done after due consideration and analysis within a systematic decision support framework with proper treatment of the associated uncertainties \citep{french1995uncertainty}. Such a framework is a defensible and powerful continuous assessment tool and allows policymakers to feed information (for example, the statistics and estimates the UK government obtains from various sources, including its Scientific Advisory Group for Emergencies (SAGE)), and their priorities into the model when it comes to decision making. At the time of writing, the COVID-19 decision support tools in the literature (discussed in Section \ref{sec:related_research}) fall short on providing such a systematic and statistically sound framework that any DC could utilise to compare their options under their self-determined constraints. However, established Bayesian methodologies have been developed over a number of decades to support DCs, capable of respecting both the preferences of the centre and the uncertainties around predictions that inform such decisions, including methods that help compare a variety of different priority weightings. This paper outlines how such a Bayesian decision analysis can help evaluate the efficacy of different COVID-19 countermeasure strategies. 

\section{Related research}
\label{sec:related_research}
\label{practical considerations}
Bayesian decision analyses have been successfully adapted and implemented for various applications, including nuclear disaster support \citep{geldermann2009multi}, food security management \citep{barons2020decision}, industrial risk management \citep{rikalovic2014gis} and environmental planning \citep{mattiussi2014decision}. For more Bayesian and frequentist applications of decision support systems, see \citet{eom2006survey}. The COVID-19 threat is fast moving, driven not only by the stochastic spread of the virus but also the dynamic control exerted by government on the activities of its agents and the general population to the unfolding crisis. Any support tool for managing this crisis needs to acknowledge that, while governments may need to switch between lockdown regimes to increase or decrease the severity of social restrictions in response to the virus prevalence, the general population may tire of the uncertainty in their lives attributable to constantly changing restrictions. In this paper, we draw from previous work \citep{leonelli2015bayesian} to demonstrate how to create a multi-attribute decision support system (DSS) under a Bayesian approach to address COVID-19 decision making challenges.

We note here that a Bayesian approach has been used for other COVID-19 studies \citep{dehning2020inferring, verma2020global, neil2020bayesian, mbuvha2020Bayesian}, although not for the purpose of providing a decision support framework. 
Finally, note that in many situations, historical data alone may be inadequate for estimating aggregate utilities from the various attributes. Continuously generated data for our recent and evolving COVID-19 situation, along with empirical data from past epidemics and pandemics are useful to a certain extent but are not sufficient. In this case, it is useful to elicit the required estimates and their associated uncertainties through discussions among a panel of domain experts. The iterative nature of such discussions and the development of the corresponding DSS are detailed in \cite{barons2018eliciting}. The structure and estimates of such a DSS would need to be repeatedly reviewed as more information comes to light (e.g. development of a vaccine or new information on immunity from the disease). Typically, the iterative improvements to the DSS are performed until it is deemed to be requisite, i.e. the DC is content that the structure of the DSS is as required \citep{phillips1984theory}. Given the urgency of the issue at hand, it is prudent to begin the process of establishing such a requisite DSS by first developing its framework to address the new challenges of the COVID-19 decision making problem. As more information and data come in, the DSS can be appropriately modified within the framework and the data can be fed into it. Thus our contribution here is the framework for a multi-attribute Bayesian DSS for COVID-19.

There have been several studies focused upon specific aspects of the COVID-19 decision-making problem such as impacts on mortality and poverty \citep{decerf2020lives}, a cost-benefit analyses of a lockdown \citep{miles2020stay, layard2020release}, and the impact of specific countermeasure strategies \citep{peto2020universal, karnon2020simple, lander2020uk}. However, none of these studies adopted a Bayesian approach which we believe is essential in this case as it not only supports uncertainty handling but also gives the government a more transparent and auditable tool. Additionally, these studies look at specific cases but do not describe their general framework. This makes it hard to adapt the existing work as new information comes in. We note here that clinical decision support tools for COVID-19 (e.g. \cite{mcrae2020clinical, wu2020development, liu2020covid, reeves2020rapid}) are beyond the scope of discussion for this paper. 

In the OR literature, Bayesian approaches have been used to model probabilities (see e.g. \cite{zafari_soyer_2020}; \cite{vargo_cogill_2015}). There have been recent calls however, for developing Bayesian subjective utility models to address deep uncertainty in evolving decision contexts (French, 2015; French, 2020). This paper addresses this by developing a framework for providing decision support in an evolving and complex decision context.

\section{A COVID-19 Decision Support Framework}
\label{sec:COVID-19DSS}
In this section, we describe how to construct a framework for a DSS to be used for examining the efficacy of different COVID-19 countermeasures within a Bayesian decision analysis. We assume that the DC satisfies the following key requirements: (1) they must agree on a single agreed rationale for their stated beliefs, and (2) the preferences and any elicited expert judgement used is adopted as their own. 

In this paper, we use the definition of a DSS from \cite{french2009decision} that, ``A decision support system is a computer-based system that supports the decision-making process, helping [DCs] to understand the problem before them and to form and explore the implications of their judgements, and hence to make a decision based upon understanding''.

It is convenient to break the process of creating a DSS down into four phases:
\\

\noindent \textbf{1) Elicitation of the class of strategies we might consider and the attributes of the utility function.}
We start by considering the types of strategy whose effectiveness we need to assess. In this framework, for the purpose of this paper, a strategy is defined by the regimes that might be imposed (such as a lockdown) and the thresholds (based on the system, which when reached leads to a switching of regimes). These strategies are designed to control the behaviour of the population and are typically tiered by the level of stringency these represent, i.e. the amount of disruption they cause to the normal life of the population.

The \emph{attributes} of the utility function are the different features of interest that each strategy affects. We define $\textbf{A}=\{A_1,A_2,\dots,A_m\}$ as the set of attributes of interest, with $a_i$ the value that attributes $A_i$. In the context of decision support for COVID-19, attributes which may be of interest include: the years of life saved across the population from avoiding infection by the virus; the years saved through timely medical examinations, for example through the detection of cancer and the encouragement to report to Accident and Emergency departments (A\&E) when exhibiting symptoms of a stroke or heart attack. A further threat to survival will be the response to increased poverty as categorised by the distribution of life expectancy given movements in the social class across the population induced by, for example, less effective education, unemployment or reduced employment activities. Note that all these attributes can be measured in units of years of life saved. On top of these, we may need to consider further attributes that define the quality of life of the population or its wellbeing, such as measures of depression and anxiety or economic hardship.
\\

\noindent\textbf{2) Elicitation of a quantification of the DC's marginal utilities of these attributes and the criterion weights.}
A subjective expected utility analysis consists of two components: a utility function $U(\boldsymbol{a})$ over a set of measured attributes $\boldsymbol{a}=\left(  a_{1},a_{2},\ldots,a_{m}\right)$ and a set of
multivariate probability densities $\left\{p_{s}(\boldsymbol{a}):s\in S\right\}$ into the future associated with each possible countermeasure strategy. Whilst the latter would be provided by the appropriate domain experts working with statisticians and mathematical modelers, the utility function $U(\boldsymbol{a})$ needs to be elicited from the DC to reflect how
they intend to frame their objectives and prioritise them. More precisely, if $\boldsymbol{a}^{-}$ and $\boldsymbol{a}^{+}$ denote what the DC perceive to be respectively the worst and best credible outcomes then for each vector of outcomes $\boldsymbol{a}$, $\left\{  \boldsymbol{a:a}^{-}%
\leq\boldsymbol{a}\leq\boldsymbol{a}^{+}\right\}  $, $U(\boldsymbol{a})$ is an increasing linear function of the probability $q(\boldsymbol{a})$ where the DC finds the outcome $\boldsymbol{a}$ with certainty equally preferable to a hypothetical situation where they are faced with obtaining the best possible outcome $\boldsymbol{a}^{+}$ with probability $q(\boldsymbol{a}%
)$ and the worst $\boldsymbol{a}^{-}$ with probability $1-q(\boldsymbol{a})$. In this way, aspects of the DC's risk aversion can be captured (see \citet{smith2010bayesian}, \citet{papamichail2003explaining}. 

On the basis of certain basic axioms (see \citet{smith2010bayesian}), various ways have been devised to indirectly elicit these preferences efficiently, effectively and specifically with less biases. One assumption that is often made, and can be checked against a DC's expressed preferences, is that the DC has value independent attributes. The attributes $\boldsymbol{a}$ are said to be \emph{value independent} if the DC always finds two strategies $s_{1}$ and $s_{2}$ (leading to densities over outcomes $\pi _{1}(\boldsymbol{a})$ and $\pi_{2}(\boldsymbol{a})$ respectively) equally preferable whenever the marginal distributions over $\pi_{1}(\boldsymbol{a})$ and $\pi_{2}(\boldsymbol{a})$ for each component attribute in $\boldsymbol{a}$ are the same. It can then be shown that we can find positive criterion weights
$\left(  k_{1},k_{2},\ldots,k_{m}\right)  $, $\sum_{i=1}^{m}k_{i}=1$ such that
$U(\boldsymbol{a})$ can be written
\[
U(\boldsymbol{a})=\sum_{i=1}^{m}k_{i}U_{i}(a_{i})
\]

where $U_{i}(a_{i})$ -- called the \emph{marginal utility} of $a_{i}$ -- is an
increasing function of its argument, and $k_{i}\geq0$, $i=1,2,\ldots,m$. Whether the attributes satisfy value independence is not an inherent feature of the attributes but is a subjective judgement of the DC. It has been found that in practice, provided the attribute vector $\boldsymbol{a}$ is carefully defined, this simple form well approximates the DC's actual utility function in the vast majority of analyses. The advantage of making this assumption is that $U(\boldsymbol{a})$ is then much easier to elicit and the output of the analysis much more transparent and easy to explain: see \citep{smith2010bayesian} for a long discussion of these points. In particular, the criterion weights of the different component attributes can be chosen to reflect their relative importance. On the other hand, the form of the marginal utilities can be chosen to reflect the extent to which a DC considers an outcome good or bad relative to its extremes for each single attribute in turn. Henceforth, in this paper, we will assume the DC has value independent attributes.
\\

\noindent \textbf{3) Building a probabilistic model of each attribute and performing a Bayesian analysis which combines available data with probabilistic expert judgements.}
This is the process of obtaining $p(a_i|s)$: the probability of attribute $i$ having value $a_i$ given strategy $s$. In the context of COVID-19, this involves obtaining a probabilistic model for the number of deaths based on the strategy taken. This requires some sophistication since the spread of the virus at any time is complex and uncertain as it is a function of both the regime that a strategy is in at that time and the latent state of the disease. Therefore, any model informing the DC needs to be able to accommodate expert judgements, epidemiological modelling and data, all synthesised through Bayesian techniques; where in complex cases these techniques would tend to be computational.
\\

\noindent\textbf{4) Calculating a final score on each attribute as an expectation of its utility function.}
These scores are weighted by the elicited criterion weight to provide expected utility scores for each of the strategies considered. Note that there is always an explanatory rationale to whether or not a strategy scores well based on each of the different individual attribute scores and the predictions of the uncertain consequences of each strategy. 

As the attributes chosen are value independent,(\cite{keeney1993decisions, insua_french_2010}), the subjective expected utility score $\overline{U}(s)$ of each strategy $s$ over the $m$ attributes $a_{1},a_{2},\ldots,a_{m}$ with respective marginal utilities $U_{1},U_{2},\ldots,U_{m}$ and the associated criterion weights $k_{1},k_{2},\ldots,k_{m}$ elicited in phase 2, is given by %

\begin{equation}
\overline{U}(s)=\sum_{i=1}^{m}k_{i}\overline{U}_{i}(s).
\label{eq:utility}    
\end{equation}
Here, the marginal expected utility scores for each attribute $j$ given strategy $s$ are given by

\begin{equation}
\overline{U}_{i}(s)=\int U_{i}(a_{i})dp(a_{i}|s)
\end{equation}

where $p(a_{i}|s)$ is the DC's subjective probability distribution built in phase 3 and $U_i(a_i)$ is the marginal utility function elicited in phase 2. Note here that in this context it is natural to define the attributes to extend to a point in time when the virus is no longer a threat. In this sense, the eventual utility scores associated with a given attribute is always uncertain throughout the decision making process.

\section{A Simplified Bayesian Analysis of COVID-19 Strategies}
\label{sec:experiments}
In this section, we give a simple illustration of how a formal Bayesian multi-criteria decision analysis, as described in Section \ref{sec:COVID-19DSS}, could have been used to help DCs weigh the efficacy of different options open to them. Typically, marginal utility functions would need to be elicited. We have mentioned above that these would typically be risk-averse, which would have the effect of reducing expected utility scores when the outcomes of associated strategies are more uncertain. However, because our main focus in this example is how the Bayesian multi-criteria decision analysis balances the efficacy achieved associated with different options, we shall in the first instance make the common assumption that each of the marginal utilities is linear in its attribute. 

By making this assumption, the analysis becomes much simpler because the DC's uncertainty only enters into calculations \textit{through the expectations of the random variables defining the future}. For simplicity, we simulate the effects of each strategy up to 40 weeks \footnote{The code used to generate this example is available here: https://github.com/peterrhysstrong/COVID-19-DSS}.

In this example we include an analysis of the effects of policies on health issues. For simplicity, we do not include the effects of the policies on the economy or the political environment. However, in a real-world analysis, these would also need to be assessed.

\subsection{Class of strategies and attributes}
\subsubsection{Regimes}
In practice, there is a large set of different regimes to consider, including potentially all possible combinations of preventative measures such as: closing schools, making mask wearing compulsory and promoting working from home. Here, we have chosen to limit our example to 3 regimes that the strategies will be able to switch between for simplicity. These loosely correspond to a regime implemented so far by the UK government. The three regimes are: 

\begin{itemize}
\item $r_{0}$ -- \textbf{No lockdown}: No restrictions. 

\item $r_{1}$ -- \textbf{Partial lockdown}: Return to work and school, non-essential businesses open with 1m+ distancing.

\item $r_{2}$ -- \textbf{Complete lockdown}: Work from home in effect, ban on non-essential interactions, complete closure of schools and non-essential businesses.

\end{itemize}

The first two regimes are analogous to the UK policies in effect on 26th March and 4th July respectively. A strategy determines when and how the transitions between these regimes occur, with the possibility of no transitions allowed.

\subsubsection{Strategies}
As the DC will be responding to the spread of the disease in a population which is susceptible to shocks such as large outbreaks, strategies considered should naturally include those which include switching between these three regimes.

In this example, we will consider several strategies that have different thresholds that give the transitions between regimes. At the time of writing, the thresholds on what measures the government was using to inform their strategies were not publicly available. In lieu of this, the transition thresholds for our example were based around the decisions made by England if they were based on the metrics defined in our example. 


In practice, a DC would likely be interested in a much wider range of strategies. However in our example, for ease of analysis, we will narrow down the set of strategies that we are interested in by putting the following conditions on them:
\begin{itemize}
    \item At most, one switch between regimes can take place in a week.
    \item Within the timescale of the simulation, once restrictions have began, we do not return to no restrictions.
    \item The transitions between regimes here are set such that they are dependent on the cumulative reported deaths, which is assumed to be known and the reported proportion of the population infected, which is assumed to be known with a delay of a week. In practice, this could be estimated from the proportion of positive tests from a sample of the population with the information being made available with a delay.
\end{itemize}

\begin{figure}[H]
\centering
 \includegraphics[width=0.5\textwidth]{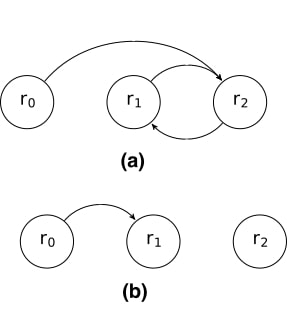}
\caption{State-space diagram of transitions between the three regimes for the two different types of strategies. (a) Strategies including lockdowns. (b) Strategies without lockdowns. }
\label{fig:exregimes}
\end{figure}

Thus, the state-space diagram for each countermeasure strategy considered corresponds to one of the figures in Figure \ref{fig:exregimes}.



 


Below, we outline the different conditions that a strategy might have for switching between regimes. \\

\noindent \textbf{Initial lockdown} ($r_0 \rightarrow r_1 / r_0 \rightarrow r_2$): Given that we are in regime $r_0$, a transition to $r_1$ or $r_2$ occurs when the cumulative deaths are larger than some chosen critical value $L$. We consider three choices of $L$:
\begin{itemize}
    \item Earlier lockdown: $L_1=100$;
    \item lockdown in line with when the UK decided to go into lockdown: $L_2=300$;
    \item Later lockdown: $L_3=500$.
\end{itemize}
Note that since the population can only transition out of $r_0$ at most once, there is no need to set out different conditions for transitioning to $r_1$ and $r_2$. Instead, each strategy specifies whether its transition out of $r_0$ is to $r_1$ or $r_2$.  \\

\noindent \textbf{Tightening of lockdown} ($r_1 \rightarrow r_2$): When a population is under regime $r_1$, a 5$\%$ rise in the observed number of cases in a week will cause a transition into $r_2$. As no tightening of lockdown had occurred at the time of this study, the choice of a 5\% rise was an arbitrary one for illustrative purposes and, in practice, a range of values would be considered. \\
 
\noindent \textbf{Easing of lockdown} ($r_2 \rightarrow r_1$): Given that a population is currently under regime $r_2$, a transition into $r_1$ occurs when the proportion of infected individuals falls below a certain critical proportion $E$ of the peak proportion of infected individuals since $r_2$ has been in effect. We consider the following choices of $E$:
 \begin{itemize}
     \item No easing: $E_0=0$;
     \item Easing in line with when the UK decided to ease lockdown: $E_1=0.12$;
     \item Quicker easing: $E_2=0.3$;
     \item Very quick easing: $E_3=0.5$.\\
 \end{itemize}

Based on these transition rules, we assume that the initial transition out of $r_0$ is to $r_2$ for all but three of the strategies we consider. A strategy denoted as $L_i\_E_j$ has initial transition from $r_0$ to $r_2$ under critical value $L_i$ and allows for easing from $r_2$ to $r_1$ under critical proportion $E_j$ where $i \in \{1,2,3\}$ and $j \in \{0,1,2,3\}$. Further, analogous to ``no complete lockdown" strategies, we define three strategies denoted by $L_i\_E*$ where the initial transition is from $r_0$ to $r_1$ under critical value $L_i$, $i \in \{1,2,3\}$. Under these three strategies, once regime $r_1$ is effected, it remains in place until the end of the simulations with no transition into $r_2$ permitted.

\subsubsection{Attributes}
\begin{figure}[H]
\centering
\begin{tikzpicture}[sibling distance=20em,
  every node/.style = {shape=rectangle, rounded corners,
    draw, align=center,
    top color=white, bottom color=blue!20}]]
  \node {Attributes}
    child { node {Short to medium term\\ $\Bar{a_1}$}
        child { node {Covid-19 Deaths\\$a_1$} }
        child { node {Delayed\\ cancer diagnoses \\$a_2$} }}
    child { node {Long term\\ $\Bar{a_2}$}
        child { node {Poverty\\$a_3$} }};
\end{tikzpicture}
\caption{Example of an attribute tree for COVID-19.}

\label{fig:atttree}
\end{figure}
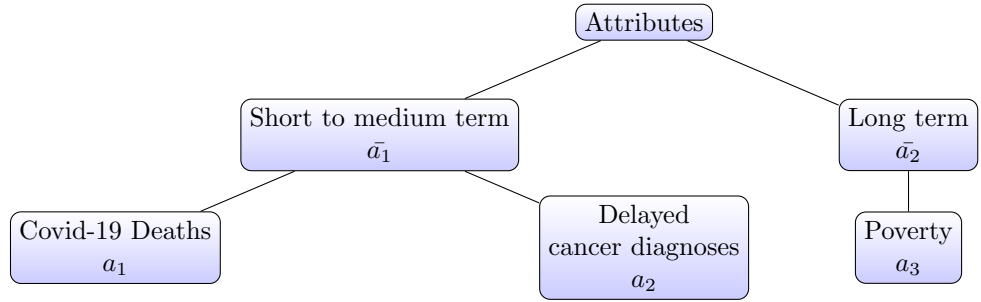

We will consider the attributes shown in Figure \ref{fig:atttree} that shows the attribute tree for our example. Since we are only considering health-related attributes, all the attributes can be measured in terms of aggregate expected life-years lost by the population. In this example, we are interested in the attributes of:
\begin{itemize}
    \item $a_1$: Life-years lost due to COVID-19.
    \item $a_2$: Excess life-years lost due to poverty.
    \item $a_3$: Excess life-years lost due to delayed cancer diagnosis.
\end{itemize}

This attribute tree can be easily adapted to add other value independent attributes (see Section \ref{sec:COVID-19DSS}) of interest for the DC such as cost of strategy implementation, economic impacts and social impacts such as quality of life. 


\subsection{Marginal utilities and criterion weights}

\subsubsection{Marginal utilities}
As in this example each of our attributes is in terms of life-years lost, we will take our marginal utility function to be the negative identity function, $U_i(a_i)=-a_i \quad \forall i$.

The values of  $a_i$ are, of course, uncertain. However, because of the assumption of preference independence and this linear utility function for the evaluation of the score, we need only elicit the subjective expectation of this quantity under each strategy. Here, again for simplicity, we have chosen to identify this expectation with the output a stochastic model of the process. We assume that the centre adopts as their expectation those delivered by relevant experts. For a theory justifying when this is an appropriate protocol for a DC to adopt, see \cite{leonelli2015bayesian} and \cite{smith2015decision}.

\subsubsection{Criterion weights}
The criterion weights for these attributes reflect how the DC prioritises them. We compare the effects of different setting of the attribute weights $k = (k_1, k_2, k_3)$ on the aggregate weighted life-years lost where weight $k_i$ is for attribute $a_i$, $i = 1,2,3$. For this example, we will consider the effect of several choices of criterion weights:
\begin{itemize}
    \item (1, 0, 0): The DC only cares about life-years lost due to COVID-19
    \item (0.5, 0.5, 0): The DC only cares about life-years lost due to COVID-19 and delayed cancer diagnosis and weights them equally.
    \item (1/3, 1/3, 1/3):The DC cares about all 3 causes of life-years lost and cares about them all equally.
    \item (0.45, 0.45, 0.1): The DC cares about all 3 causes of life-years lost and but life-years lost caused by short and medium term attributes are more important.
\end{itemize}

\subsection{Probabilistic model of each attribute}
For each attribute, we will give a probabilistic model giving us the probability of an attribute obtaining each value for each strategy, $p(a_i|s)$.

\subsubsection{COVID-19 deaths ($a_1$)}
To estimate the deaths from COVID-19 and the proportion of infected individuals in the population, we use a simple discrete-time SIRD compartmental epidemiological model. The states in this model are ``Susceptible" (S), ``Infectious" (I), ``Recovered" (R) and ``Dead" (D). The dynamics of this model can be seen in Figure \ref{sird} and are described in the supplementary material S1. Clearly, D is an absorbing state and we assume that those that have recovered from COVID-19 are immune to it. We stratify our population in this model by region and age.


\begin{figure}[H]
    \centering
    \includegraphics[width=0.8\textwidth, trim={0cm 2cm 0 0cm}] {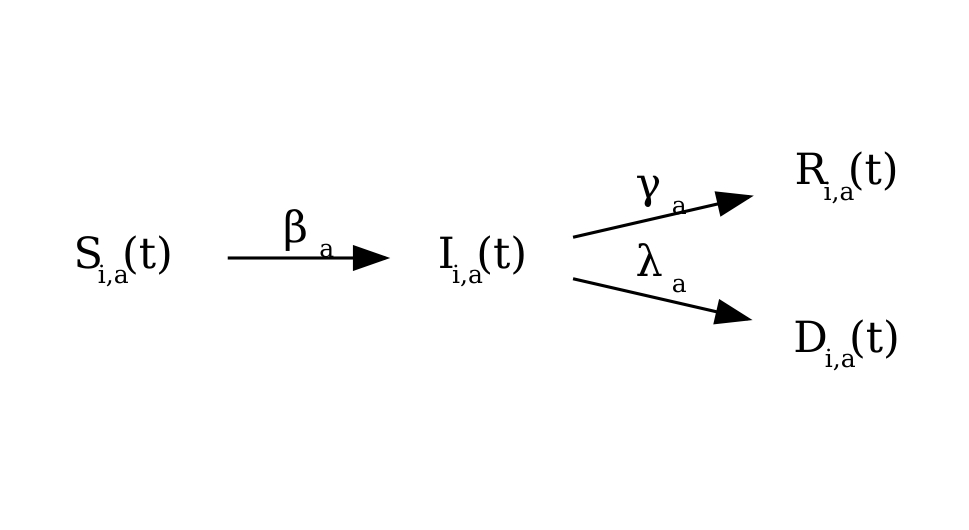}
    \caption{SIRD disease dynamics}
    \label{sird}
\end{figure}

\begin{figure}[H]
\centering
\centering
   \includegraphics[width=\textwidth]{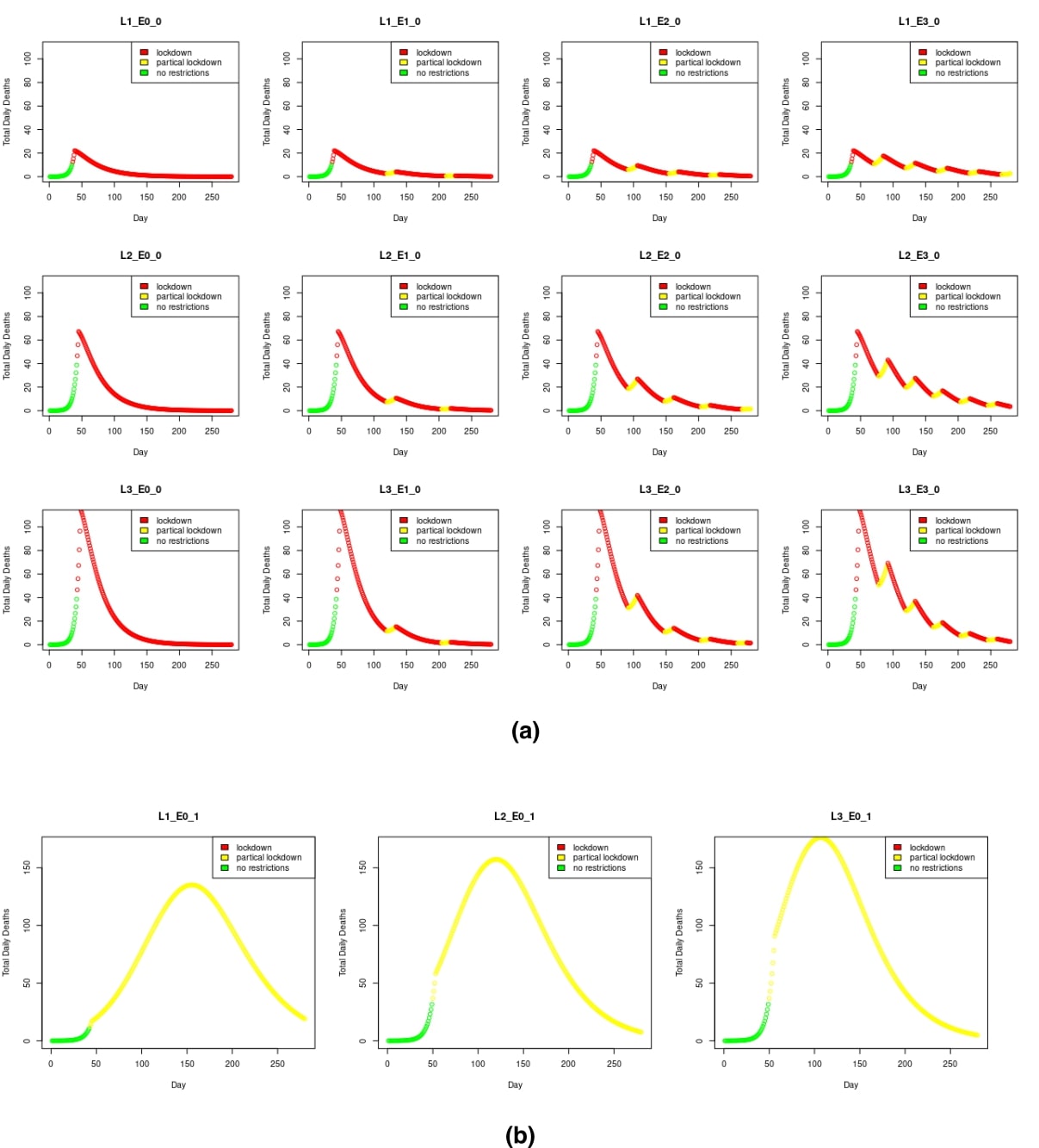}
\caption{Plots showing how the daily deaths evolve over time for one simulation within the stochastic model.(a) Strategies including lockdowns. (b) Strategies without lockdowns.}
\label{deaths_evolve}
\end{figure}

In order to capture the uncertain nature of the future, we have modelled uncertainty around the rate of infection with detail given in supplementary material S1.2. By running our stochastic simulation 1000 times we obtained expected values for both the attribute of number of deaths due to COVID-19 and the number of weeks spent in each regime. One of these simulations for each strategy is shown in Figure \ref{deaths_evolve}. Note here that in practice the expected numbers used are those provided by epidemiologists using more sophisticated models. 


\subsubsection{Delayed cancer diagnoses ($a_2$)}
In this example, deaths due to delayed diagnoses of cancer are used here as a proxy for deaths that would not have normally occurred, as a composite of missed treatments and patients who have not presented in hospital. This was chosen as a proxy as information was easily available. In practice, this could be improved on by either creating attributes for other excess deaths due to missed treatments and eliciting models of each attribute or, when that is not feasible, updating the utility function for this attribute. The latter could be done simply by assuming that deaths due to delayed cancer diagnosis will form a certain proportion of all of the deaths due to missed treatment. Estimates like this are often necessary in the early stages of a decision support system for an emergency response when there is sparse information about the underlying science.

\citet{sud2020effect} evaluate how delays to cancer diagnoses due to the COVID-19 outbreak impact survival and life-years lost. This study gives a nearly linear relationship for delay in cancer diagnoses to extra deaths per age group. Using this relationship, we are able to produce plausible values for the expected total number of deaths arising from this delay. These can be used as benchmark values the DC might use for its subjective expectations of this attribute.

As at the time of this report there is little research done on the impact of partial lockdown on cancer referrals, we will assume that the number of deaths in a partial lockdown is half of those in a total lockdown. In practice, the DC would elicit expert judgement about this value where possible and perform a sensitivity analysis.

\subsubsection{Poverty ($a_3$)}

Here, we notice that the centre has other experts available who can assess the impact on years of life lost on poverty induced by a strategy.

For example, \citet{decerf2020lives} evaluate the impact of the COVID-19 outbreak in terms of number of poverty-years. Poverty-years are calculated using COVID-19's predicted effect on GDP and estimates the number of additional people in poverty as a result (in the UK, 4.37 million).  \citet{decerf2020lives} conservatively assumes that these individuals will remain poor only for a single year.
We use \citet{decerf2020lives}'s calculation that 8.8 poverty-years equate to 1 life-year to inform our DSS, which gives us an estimate for the impact of lockdown on life-years lost due to poverty. Further information is given in the supplementary material section S2.

\subsection{Final scores calculated}
 We can now combine this all together to get Equation \ref{eq:raw_attribute} and assess which strategy maximises the utility score given the criterion weights.
 
\begin{equation}
\overline{U}(s)=\sum_{i=1}^{3}k_{i}\overline{U}_{i}(s)=\sum_{i=1}^{3}k_{i} (-\overline{a_i(s)})
    \label{eq:raw_attribute}
\end{equation}

Here $\overline{a_i(s)}$ is the expected value of attribute $i$ under strategy $s$. In this example, this is the magnitude of the utility function, $\overline{U}_{i}(s)$.

\subsection{Results}


\begin{figure}[H]
\centering
    \includegraphics[width=\textwidth]{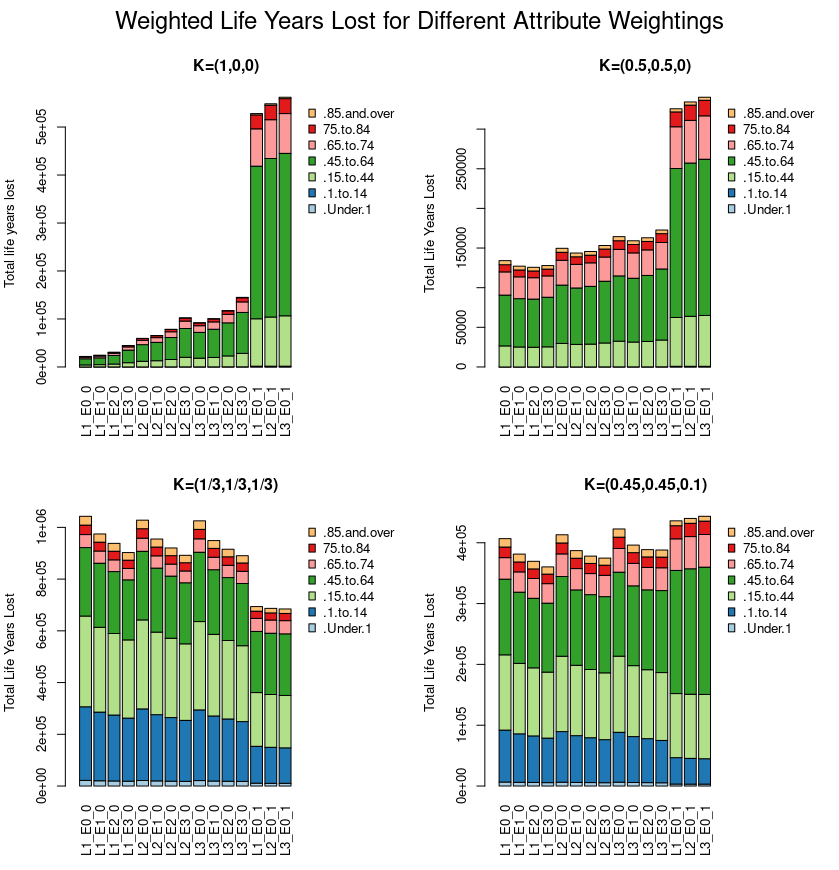}
    \caption{Aggregate weighted life-years lost under different attribute weightings for each strategy. Top left: COVID-19 deaths only. Top right: COVID-19 and delayed cancer diagnoses deaths equally weighted. Bottom left: COVID-19, delayed cancer diagnoses and poverty deaths equally weighted. Bottom right: COVID-19, delayed cancer diagnoses and poverty deaths custom weighting.}\label{attweight}
\end{figure}

From Figure \ref{attweight}, we can see that under $k = (1, 0, 0)$ where only life-years lost due to COVID-19 are of interest, strategies that do not involve a lockdown perform orders of magnitudes worse. Looking at strategies that do involve lockdown, ``earlier lockdown" and ``slower easing of lockdown" lead to fewer life-years lost. The second scenario under $k = (0.5, 0.5, 0)$, in which life-years lost due to delayed cancer diagnoses are also considered, we can begin to see the drawbacks of the lockdown with reduced disparity between strategies that include lockdown and those that do not. However, the lockdown strategies remain more desirable. This example also shows little difference between the lockdown strategies that only differed in their easing of lockdown with more significant differences depending on when the lockdown conditions were implemented.

Under $k = (1/3, 1/3, 1/3)$, which includes life-years lost due to poverty with equal weighting for all attributes, the non-lockdown strategies perform the best. It is interesting to see how small the weight on our long term impacts on health might need to be before we impose the most stringent of lockdown strategies. For  $k=(0.45,0.45,0.1)$, we treat short and medium term years lost as more important than long term years lost. While lockdown strategies generally still perform better, the disparity among all strategies is much less.


\begin{figure}[H]
\centering
    \includegraphics[width=1\textwidth]{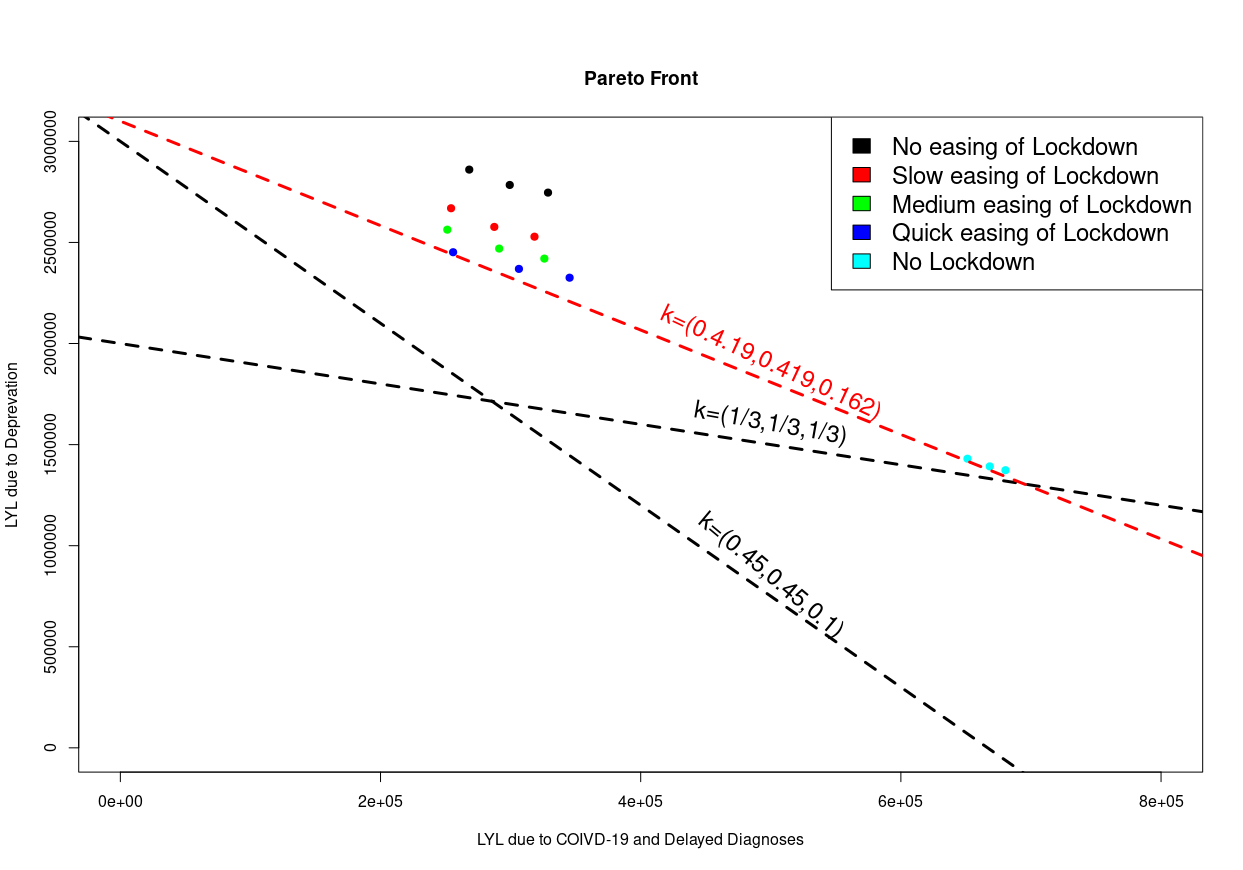}
    \caption{Pareto plots showing the trade-off between life-years lost (LYL) due to poverty (deprivation) on one hand and COVID-19 and delayed cancer diagnoses on the other. The dashed lines represent examples of different attribute weightings.}
    \label{paretototal}
\end{figure}

We further explore these trade-offs in Figure \ref{paretototal}, which  shows a Pareto front plotting life-years lost due to poverty against those lost from COVID-19 and cancer combined, assuming equal weighting. We can see how the trade-off in weights leads to different strategies being the most effective, with the weights giving the gradient of the line. This plot can help to reduce the number of strategies considered, as in this example strategies with no easing of lockdown are strictly dominated by strategies with quick easing of lockdown. Thus, there is no weighting of attributes in which we would rather have no easing. We can also find the value $c$ in $k^*=(c,c,1-2c)$ for which the DC would swap preferences between strategies that involve lockdown and those that don't. In this example, that point is at $c=0.4.19$, where short and medium term attributes are weighted as 2.58 times more important than long term attributes. At $c \geq 0.4.19$, the optimal weighting would be strategies that involve lockdown. For comparing trade-offs between more attributes at the same time, an n-dimensional Pareto front with each axis on its own label should be used. Further analyses of the trade-offs between the attributes using Pareto fronts is given in the supplementary materials S3-S4.

\section{Discussion}
\label{sec:discussion}


In this paper, we presented how a DC can create a DSS under which countermeasure strategies for COVID-19 could be evaluated in real-time, using the best expert judgements associated with various strategies that it might be feasible for the DC to enact. This DSS enables the DC to input the attributes they consider to be vital, with attribute weightings reflecting their priorities, to evaluate and choose an appropriate metric by which these attributes can be measured.

The novelty of the framework lies in demonstrating how to provide multi-attribute Bayesian decision support in evolving decision contexts. We have shown that Bayesian decision analysis is an appropriate and holistic approach for addressing uncertainty in dynamic environments as we move from short-term to longer-term considerations. Indeed, the results of a Bayesian subjective utility approach are informative and help decision makers devise action plans \citep{french_2020}.

In the results from our example, we observed the effect of different criterion weights demonstrating their effect on the optimal decision. For illustrative purposes, the analysis presented here was simplified. In order to be operationalised, it would need to be refined with additional attributes as well as a utility function and attribute weightings that reflect the priorities of the DC. Firstly, more attributes -- such as hospital admissions, well-being,  economic viability and public acceptability -- could be considered.

Further, such a DSS could be extended by considering the impact of other countries' management of COVID-19, caused by a spillover effect. For example, the UK's economy is likely to be affected by COVID-19 spreading in other countries regardless of the UK's state of lockdown. Our model implicitly assumes that such further causes of deprivation are additive, whereas the relationship between impacts to domestic employment and international economic variables on the differential scores associated with different options is likely to be more subtle. However, we note that, by embarking on this decision analysis, we are drawn into trying to quantify such issues and to fold these important considerations into the analysis; this DSS gives a framework for introducing such studies as these become available.

Another critical issue that could be incorporated would be the capacity of the health provider, e.g. the National Health Service (NHS) in the UK, both with respect to the number of available beds and the number of doctors and nurses available to treat COVID-19 as well as non-COVID-19 patients. Hospitals functioning close to full capacity may lead to increase in the mortality rate of hospitalised COVID-19 patients \citep{wilde2021association} as well as temporary closure of all non-urgent hospital procedures \citep{covidsurg2020global}. Thus, if the NHS were to be overwhelmed, the underlying model would need to be adapted to reflect the increased life years lost due to the added pressure. Indeed, in the UK, COVID-19 policy decisions have been influenced by the capacity of the NHS \citep{guardian_nhs_overwhelm}, and measures such as introducing additional capacity through construction of temporary hospital facilities (known in the UK as Nightingale Hospitals) \citep{nightingale_nhs} were effected to prevent overwhelming the NHS. These capacities could be incorporated (1) in deciding the thresholds for changes in regimes and (2) by choosing hospital admissions as an attribute. The utility function of this attribute could be an indicator function that indicates when the hospital capacity is exceeded. Assigning a very large criterion weight to such an attribute would have the effect of making any strategy that would lead to the threshold being passed strictly dominated by other strategies and therefore, not considered.

We also assumed that all the marginal utilities were linear. In contexts like the one above, the DC tends to be risk-averse. The mathematics means that the scores we assign to different options which lead to more uncertain outcomes are penalised. There are two elements that come into play here. The first issue is that the impacts of previously untried strategies are likely to be more uncertain. A risk-averse marginal utility would therefore tend to down-weight the scores of less well-tried options. The second is that attributes which are intrinsically more uncertain will tend to be given less weight across all strategies. For example, the implicit assumption of giving negligible weight to non-poverty related economic effects could be justified if the differential economic effects of one strategy against another were extremely uncertain. However, once the marginal utilities have been elicited, the framework is able to score the different options in light of this and the type of analysis above can be adapted. Further, compliance of the population to flip-flopping between regimes could be accounted for by considering it within the probabilistic predictions of the attributes.

Finally, we have used a na{\"i}ve predictive model of the progress of the disease. This dynamic model could obviously be improved; see \citet{keeling2021predictions, aguilar2020investigating}. In particular, we could choose a stochastic disease model such as \citet{abrams2020modeling}. Assumptions on immunity after recovery might also need to be revised as relevant studies become available; for example \citet{long2020clinical, liu2020prevalence}. 

In conclusion, further research is needed to extend on this work in all the ways mentioned above. However, we have shown, using as simple example, how a DSS could be created and analysed to assess different countermeasure strategies for COVID-19. We have also shown how this could be done in a real time setting -- where there is only basic scientific understanding of such a disease and the impacts of its countermeasures, as witnessed for COVID-19 -- where DCs may need to rely on only coarse descriptions given by the experts when many judgements are unknown or unsubstantiated by evidence.

\printbibliography[title = References]

\newpage
\begin{appendices}

\section*{S1: Disease dynamic model}
\label{ddm supp}
In our example, the UK will be split up into 14 regions and the 7 age groups used by the ONS, each with population $N_{i}$ which we will assume to be constant. Table \ref{sup:tab:regions_ages} shows the regions and age groups. We further split up these populations into age groups giving us $N_{i,a}$. This model has the constraint:
\begin{equation}
    S_{i,a}(t)+I_{i,a}(t)+R_{i,a}(t)+D_{i,a}(t)=N_{i,a}
    \label{N}
\end{equation}
where $i= 1,2,\dots,14$ are the locations and $a=1,2, \dots,7$ are the age classifications. For any time $t$, the sum of individuals in each state in location $i$ in age group $a$ is equal to the number of individuals in the sub population $i$ in age group $a$. The disease dynamics are shown in Equations (\ref{N}-\ref{D}).

\begin{equation}
\Delta S_{i,a}(t) = - \beta_{a} \frac{I_{i}(t)}{N_i}S_{i,a}(t)  
\label{S}
\end{equation}

\begin{equation}
\Delta I_{i,a}(t) =  \beta_{a} \frac{I_{i}(t)}{N_i}S_{i,a}(t)-(\gamma_a+\lambda_a)I_{i,a}(t)   
\label{I}
\end{equation}

\begin{equation}
\Delta R_{i,a}(t) = \gamma_a I_{i,a}(t)
\label{R}
\end{equation}

\begin{equation}
\Delta D_{i,a}(t) = \lambda_a I_{i,a}(t)
\label{D}
\end{equation}
Each time step in this model represents a day, with $\beta_{a}$, $\gamma_a$ and $\lambda_a$ being the retrospective daily rate of infection, recovery and death. We will assume each of these rates are independent of location and time but are age-dependent (hence, these are indexed only by age classification $a$ and not indexed by location $i$ or time $t$). $\beta_{a}$ is dependent on what strategy is currently being enforced to fight the infection. $I_i(t)=\sum_a I_{i,a}(t)$ is the total number of infectious people in location $i$. This is used in Equations \ref{S} and \ref{I} as we are assuming uniform mixing between age groups.

\begin{table}[H]
\centering
\begin{tabular}{|c|c|}
\hline
\textbf{Regions}         & \textbf{Age groups} \\ \hline
North East               & Under 1                                 \\
North West              & 1-14                                    \\
Yorkshire and the Humber & 15-44                                   \\
East Midlands           & 45-64                                   \\ 
West Midlands                                  & 65-74                                   \\
East                                           & 75-84                                   \\
London                                         & 85 and over                             \\
South East                                     &                                         \\
South West                                     &                                         \\
Wales                                          &                                         \\
Scotland                                       &                                         \\ 
Northern Ireland                               &                      \\\hline                  
\end{tabular}
\caption{Table showing regions and age groups used in our analysis}
\label{sup:tab:regions_ages}
\end{table}

\subsubsection*{S1.1: $\lambda_a$ and $\gamma_a$}
To estimate $\gamma_a$ and $\lambda_a$, we will use estimates of infection fatality ratios in each age group $D_a$ from \cite{ferguson2020report}. We will assume that 
28 days is sufficient time after a positive test for an individual's death not to be thought of as due to coronavirus. We will therefore model that after 28 days, the probability of still being infected is 0.05. Using this, we can estimate $\gamma_a$ and $\lambda_a$ by solving the following:
\begin{equation}
    \sum_{t=0}^{27} (\gamma_a+\lambda_a)(1-\gamma_a-\lambda_a)^t=0.95
    \label{notI}
\end{equation}
\begin{equation}
    \sum_{t=0}^{\infty} \lambda_a(1-\gamma_a-\lambda_a)^t=D_a
    \label{Dead}
\end{equation}
Equations \ref{notI} and \ref{Dead} give the probability of either recovering or dying in 28 days since infection and the probability of dying after becoming infected respectively. $(1-\gamma_a-\lambda_a)^t$ is the probability that an infected individual of age $a$ is still infectious at time step $t$. 

Equation \ref{notI} can be solved algebraically using the geometric series formula, giving $\lambda_a+\gamma_a=0.101466$. Combining this with Equation \ref{Dead} can be used to give each $\lambda_a$ and $\gamma_a$ values.
\subsubsection*{S1.2: $\beta_{a}$}
We can calculate an estimate for $\beta_{a}$ by splitting it into $p$ (the probability of becoming infected after contact with an infected individual) multiplied by $c_{a}$ (the average number of contacts an individual in age group $a$ has per day while strategy $s$ is being enforced). Under normal conditions, the average number of contacts for individuals of each age can be found from \cite{mossong2008social}. We assume under partial lockdown that the number of contacts will be half of this and within lockdown individuals of any age group will have 3 contacts. We can work out an estimate of $p$ for our example by considering the value of $R_0$, the average number of people an infectious person will infect in a fully susceptible population. There have been many estimates of $R_0$ for COVID-19.  \cite{liu2020reproductive} gives a review of these and gives a median value of 2.79. We can calculate $p$ as follows:
\begin{equation}
    R_0=\frac{\sum_{a \in A}n_a\sum_{t=0}^{\infty}c_a p(1-\gamma_a-\lambda_a)^t}{\sum_{a \in A}n_a}
\end{equation}
Here $\sum_{t=0}^{\infty}c_a p(1-\gamma_a-\lambda_a)^t$ is the average number of infections that an infectious individual in age group $a$ will cause. The population weighted average of this gives $R_0$.
This can be rewritten to give $p$ and solved using the previously described values.
\begin{equation}
    p=\frac{R_0\sum_{a \in A}n_a}{\sum_{a \in A}\sum_{t=0}^{\inf}n_ac_a (1-\gamma_a-\lambda_a)^t}
\end{equation}
This gives $p$=0.023.

To add stochasticity into our model, we ran the simulated SIRD model 1000 times with $p$ distributed from a log normal with mean 0.023 and 0.1 standard deviation on the log scale.

\subsubsection*{S1.3: Initial conditions}
The model was initialised on the 2nd March: the first date when all regions in the UK had at least one coronavirus case.

\newpage
\section*{S2: Other attributes} \label{other attributes supp}
\subsection*{S2.1: Delayed cancer diagnoses} 
\cite{sud2020effect} use 10-year age- and stage-stratified cancer survival estimates in England for 20 different common tumour types present in adults using information from Public Health England. Using these estimates and annual numbers of early-stage cancers diagnosed via the `2-week-wait’ pathway, they calculate probable lives and life-years lost in England dependant on the length of a lockdown and how many symptomatic patients’ assessments are delayed by lockdown.

The research primarily focuses on early stage I-III cancers; they assume delays to stage IV cancer treatment will not affect 10 year survival. They also focus on diagnosis delays instead of treatment delays; it is assumed that there is no delay to treatment, which the authors accept is unlikely in reality. The authors calculate the attributable lives lost by length of delay due to the pausing of the `2-week-wait’ pathway for 1-6 months for 20 different tumour types. They also age-stratify each cancer type to denote the proportion of tumours prevalent in each age group. By combining this information, we can identify how many lives are lost for each cancer in each age group dependant on how long the `2-week-wait’ pathway is not operating at 100\% capacity. 

\subsection*{S2.2: Deaths due to poverty}
\citet{decerf2020lives} estimates the number of additional poverty years for the UK at 4,370,000 and estimate that 8.8 poverty years is equivalent 1 life-year lost. In Table \ref{tab:poverty_estimates}, we use the estimates of the population living under the poverty threshold in the UK, stratified by age (`children', `working-age' and `pension-age') as given in \citet{social2019} and assume that the additional poverty years are distributed the same way.

\begin{table}[hb]
\begin{tabular}{c|c|c|cl}
Age Group & \begin{tabular}[c]{@{}c@{}}2017/18 Estimate of number \\ of people in poverty\end{tabular} & \begin{tabular}[c]{@{}c@{}}Additional \\ poverty years\end{tabular} & \begin{tabular}[c]{@{}c@{}}Equivalent \\ life-years lost\end{tabular} &  \\ \cline{1-4}
Children & 4.6 & 1.41 & 0.16 &  \\
Working-age & 8.3 & 2.54 & 0.29 &  \\
Pension-age & 1.3 & 0.42 & 0.048 &  \\ \cline{1-4}
Total & 14.2 & 4.37 & 0.498 & 
\end{tabular}

    \caption{Additional Poverty years in 2020 in the UK. All estimates are in millions.}
    \label{tab:poverty_estimates}
\end{table}

We assume that children comprise of those under 16 years of age, working-age as those between 16 and 65 years of age, and pension-age as those over 65 years. \citet{decerf2020lives} evaluated the effect of the COVID-19 pandemic on poverty years as of early June 2020. For the UK, this period corresponds to the time that the UK was more or less under complete lockdown. In our paper, we shall assume -- for illustrative purposes -- that the impacts on poverty years were caused as a direct effect of the lockdown and whose alleviation would be proportional to the removal of lockdown restrictions. As with cancer deaths, at the time of this report there is little research done on the impact of partial lockdown on poverty and therefore we will assume that the number of additional years in poverty as a result of a partial lockdown is half of those in a total lockdown. 

\cite{decerf2020lives} state that their methodology gives a conservative estimate by providing a lower bound for additional years spent in poverty due to COVID-19 countermeasure strategies. They make the assumption that the condition of individuals who are faced with poverty will improve within a single year and will not have further lasting effects. Thus, the premature mortality effects of poverty are not actually considered \citep{baland2019too}. Additionally, it is assumed that the GDP reduction caused by the economic contraction is distribution neutral, i.e. the inequality in the society remains constant. However, this is known not to be the case \citep{van2020covid}. While there are several shortcomings to these estimates, we use them in our paper for illustrative purposes and for the lack of better estimates regarding deprivation caused by COVID-19 and the lockdown at this point of the pandemic \citep{miles2020stay}.

\subsection*{S2.3: Age-stratification}
Occasionally, the age-stratification from our data sources is not the same as the age-stratification in this study. Where possible, the age-stratifications were combined. Where this was not possible, the group was split proportionally; for example, the 60-69 age group in \cite{sud2020effect} was split 50/50 into the appropriate age ranges for this research of 45-64 and 65-74. This assumes that the proportion of cases is uniformly distributed across the age group.
\newpage

\section*{S3: Life-years lost under each attribute}
\begin{sidewaysfigure}
    \includegraphics[width=\textwidth]{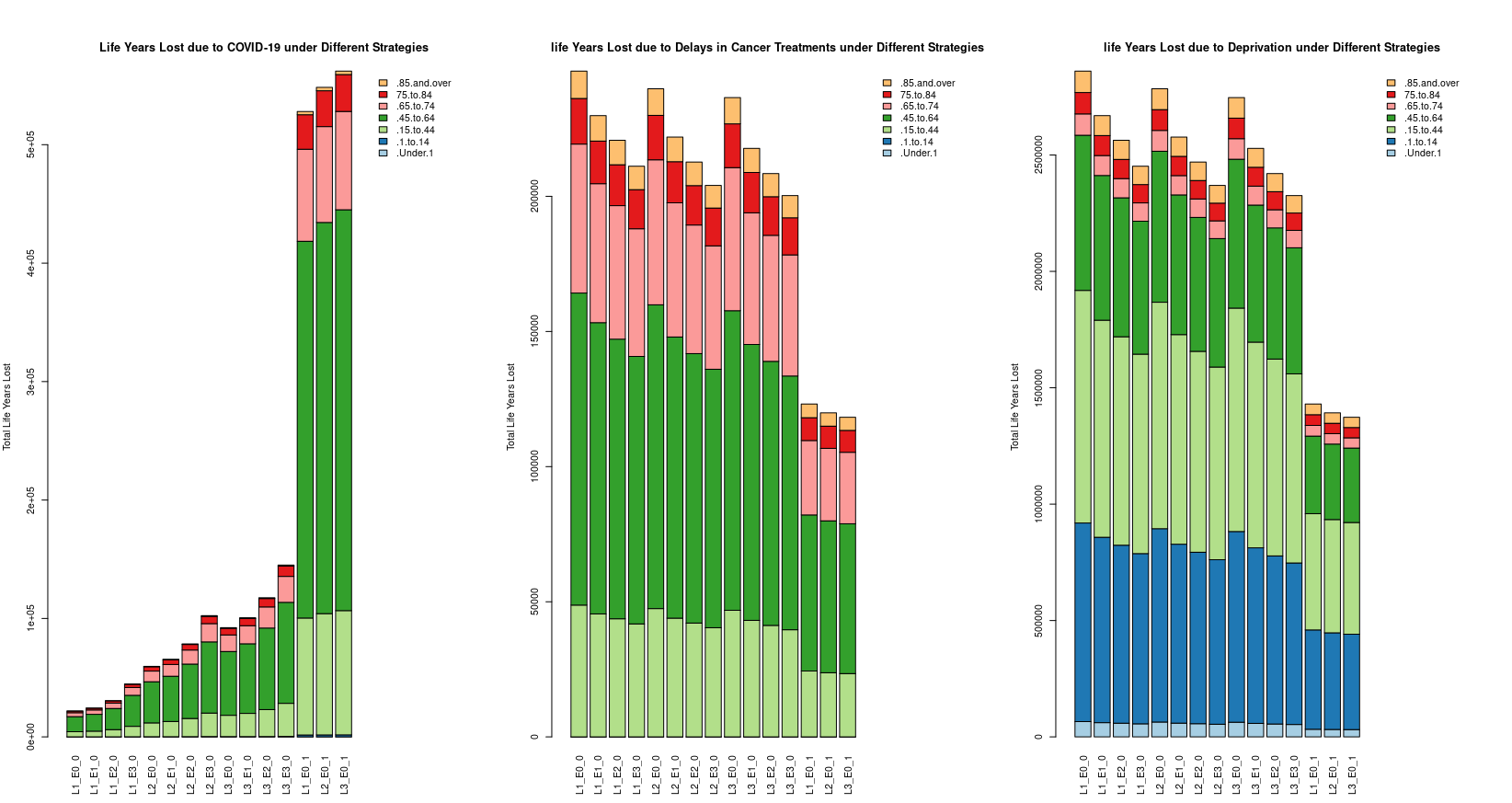}
    \caption{Diagram showing the Life-Years Lost under each attribute due to each strategy.}
\end{sidewaysfigure}

\newpage

\section*{S4: Age comparison of strategies}

\begin{figure}[H]%
    \centering
    \centering
    \includegraphics[width=10cm]{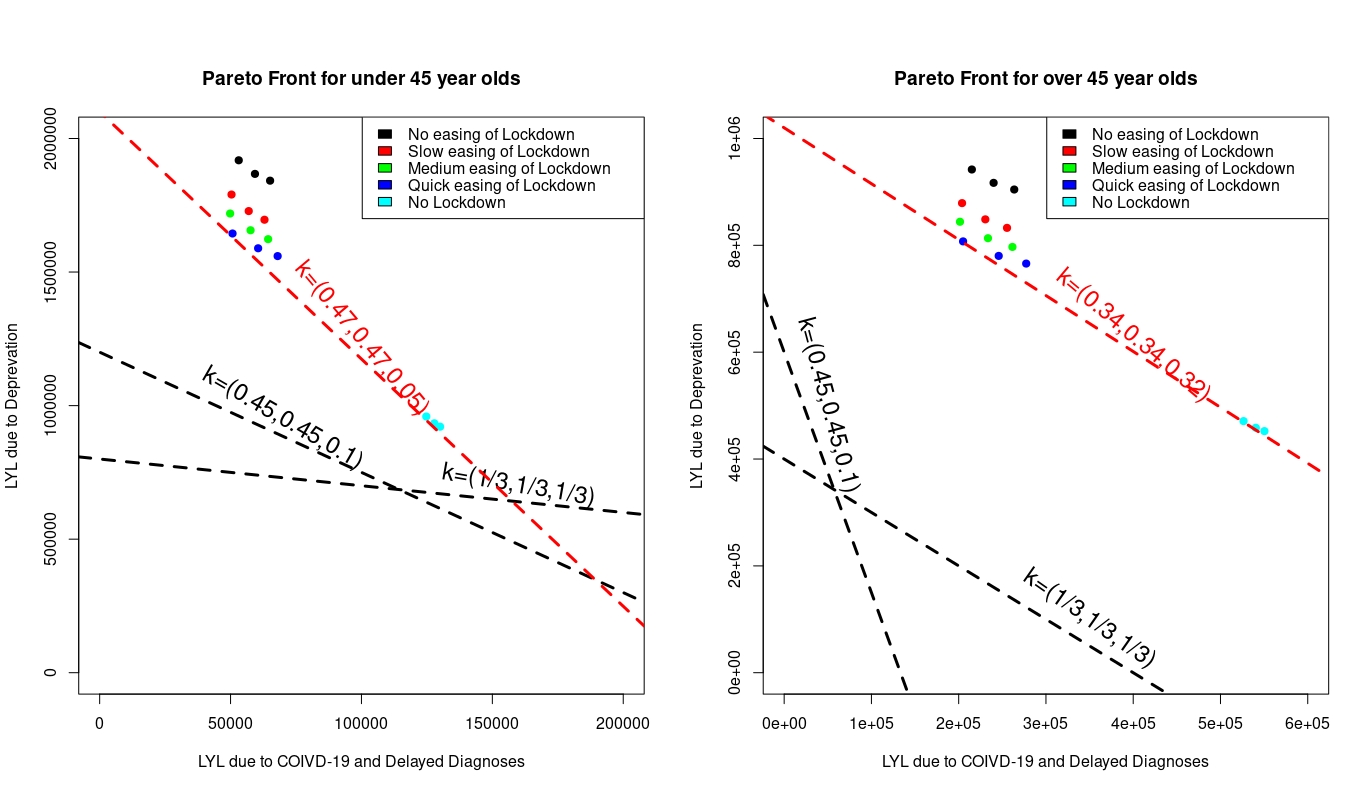}
    \caption{Pareto plots showing the trade-offs of life-years lost due to COVID-19 and delayed cancer diagnoses to life-years lost due to poverty (deprivation) comparing the effects of different trade-offs on the two age classifications of over and under 45 year olds. The dashed lines represent examples of different attribute weightings with the one in red representing the critical weighting between lockdown and non lockdown strategies.}%
    \label{age_parate}%
\end{figure}

We explore how the strategy impacts different age groups in Figure \ref{age_parate}. We look at the critical weightings $k^*_o=(c_o,c_o,(1-2c_o))$ and $k^*_u=(c_u,c_u,(1-2c_u))$ for switching between strategies that involve lockdown and those that don't for groups of over and under 45 years old respectively. We get $c_u= 0.474$ and $c_o=0.338$, which means that when selecting strategies that focus on people over the age of 45, you would have to value short and medium term attributes as 1.05 times as important than those strategies that do not involve lockdown. When optimising to benefit people under the age of 45, you would have to value short and medium term attributes as 9.26 times as important to consider strategies than those which involve lockdowns.



\end{appendices}

\end{document}